\documentclass[preprint2]{aastex}

\newcommand{\etal}{{\em et al.}}
\newcommand{\beq}{\begin{equation}}
\newcommand{\eeq}{\end{equation}}
\newcommand{\tnm}{\tablenotemark}
\newcommand{\aam}{\altaffilmark}
\newcommand{\HI}{H\mbox{\sc\small ~I~}}
\newcommand{\HII}{H\mbox{\sc\small ~II~}}
\newcommand{\kms}{km s$^{-1}$}
\newcommand{\mas}{milliarcsecond}
\newcommand{\thetavec}{\mbox{\boldmath $\theta$}}
\newcommand{\dtd}{\Delta t_{\rm d}}
\newcommand{\dnud}{\Delta\nu_{\rm d}}
\newcommand{\Viss}{V_{\rm ISS}}
\newcommand{\Vissu}{V_{\rm ISS,5/3,u}} 
\newcommand{\Aissu}{A_{\rm ISS, 5/3, u}}
\newcommand{\Wc}{W_{\rm C}}
\newcommand{\Wdpm}{W_{\rm D,PM}}
\newcommand{\Vpperp}{{V_p}_{\perp}}
\newcommand{\Cnsqr}{C_n^2}
\newcommand{\Cnsqrtc}{C_{n,{\rm TC93}}^{2}}
\newcommand{\Cnsqrs}{C_{n,s}^2}
\newcommand{\dSM}{\Delta\mbox{SM}}
\newcommand{\dDM}{\Delta\mbox{DM}}
\newcommand{\Ds}{\Delta s}

\slugcomment{ApJ submitted, 3 October 2000}

\begin{document}
\title{Parallax and Kinematics of PSR B0919+06 
       from VLBA Astrometry and Interstellar Scintillometry}
\shorttitle{Parallax and Kinematics of PSR B0919+06}

\author{S.~Chatterjee\aam{1}, J.~M.~Cordes\aam{1},
T.~J.~W.~Lazio\aam{2}, W.~M.~Goss\aam{3}, E.~B.~Fomalont\aam{4},
J.~M.~Benson\aam{3}}   
\altaffiltext{1}{Cornell University, Ithaca, NY 14853 \\
Contact email: shami@astro.cornell.edu}
\altaffiltext{2}{Naval Research Laboratory, Washington, DC 20375}
\altaffiltext{3}{NRAO, Socorro, NM 87801}
\altaffiltext{4}{NRAO, Charlottesville, VA 22903}

\begin{abstract}

Results are presented from a long-term astrometry program on PSR
B0919+06 using the NRAO Very Long Baseline Array. With ten
observations (seven epochs) between 1994--2000, we measure a proper
motion $\mu_{\alpha} = 18.35 \pm 0.06$ mas yr$^{-1}$, $\mu_{\delta} =
86.56 \pm 0.12$ mas yr$^{-1}$, and a parallax $\pi = 0.83 \pm 0.13$
mas (68\% confidence intervals). This yields a pulsar distance of
$1.21 \pm 0.19$ kpc, making PSR B0919+06 the farthest pulsar for which
a trigonometric parallax has been obtained, and the implied pulsar
transverse speed is $505 \pm 80$ \kms. Combining the distance estimate
with interstellar scintillation data spanning 20 years, we infer the
existence of a patchy or clumpy scattering screen along the line of
sight in addition to the distributed electron density predicted by
models for the Galaxy, and constrain the location of this scattering
region to within $\sim 250$ parsecs of the Sun.  Comparison with the
lines of sight towards other pulsars in the same quadrant of the
Galaxy permits refinement of our knowledge of the local interstellar
matter in this direction.  \\
\end{abstract}


\section{Introduction}

Since the discovery of pulsars, it has been recognized that the
dispersion and scattering of pulsar signals provide unique information
about the intervening medium. The distance to most pulsars is
estimated using the observed dispersion measure (DM) and a model for
the Galactic electron density distribution \citep[e.g.][hereafter
TC93]{TC93}. Where available, a model-independent distance from annual
trigonometric parallax provides crucial calibration information for
this model, as well as allowing the absolute luminosity of the pulsar
to be derived.

Pulsar astrometry also provides the observational evidence required to
investigate several other questions. Proper motion measurements
(especially in conjunction with reliable distance estimates) allow
verification of pulsar--supernova remnant associations \citep[e.g.][]{K98}. 
The inferred speeds of pulsars constrain the minimum
asymmetry in supernova core-collapse processes, or other sources of
kick velocities. Pulsar population statistics, selection-effect
biases, and planetary--extragalactic reference frame ties have also
been addressed by astrometry \citep[e.g.,][]{LL94,CC98,BCR+96}.

At present, there are only a handful of model-independent distances to
pulsars. In a recent summary, \citet{TBM+99} list nine such objects. In
\S\ref{Sec:VLBI} of this paper, we present results from a long-term
astrometry program on PSR B0919+06 using the NRAO Very Long Baseline
Array (VLBA). We measure the proper motion and trigonometric parallax
for this pulsar, and infer a distance of $\sim 1.2$ kpc and a
transverse speed of $\sim 500$ \kms, comparable to the mean population
speed.

PSR B0919+06 has a DM of 27.31 pc cm$^{-3}$ \citep{PW92}, yielding a
mean electron density of 0.023 cm$^{-3}$ for the interstellar medium
(ISM) along this line of sight. However, interstellar scintillation
(ISS) and scattering of the pulsar signals can be used to extract much
more information about the line of sight than simply this mean
electron density. In \S\ref{Sec:ISS}, we review the necessary
formalism, and analyze the published data spanning $\sim 20$ years for
this pulsar. Our analysis combines data from very long baseline
interferometry (VLBI) and ISS, in order to infer the distribution of
scattering material along the line of sight. We find the need for
extra scattering material in addition to the TC93 distribution; with
the assumption that this material is present in the form of a screen
or clump with variable scattering strength, we constrain the scren
location to the local interstellar medium, within $\sim 250$ parsecs
of the Sun.

This analysis, along with DM and distance measurements for other
pulsars in the third quadrant of our Galaxy, allows refinement of our
knowledge of the local electron density distribution. Besides being a
situation of interest in its own right, this analysis also serves as a
model for the hybrid VLBI--ISS technique applied to the local ISM.

\section{Obtaining a VLBA Proper Motion and Parallax\label{Sec:VLBI}}

We undertook observations of PSR B0919+06 using the VLBA, over seven
epochs (ten observations) from 1994 to 2000. The details of data
acquisition, processing and calibration have been discussed by
\citet{FGBC99}, who presented a preliminary result based on the first
four epochs of this dataset. For completeness, we summarize the
data reduction procedure here.

\subsection{Phase Referencing and In-Beam Calibration}

In VLBI observations, the visibility phase has to be estimated for
each baseline (``fringe fitting'') in order to correct for variations
in clock offsets, station positions and atmospheric propagation
effects.  However, fringe fitting is not possible for weak sources,
which lack sufficient signal-to-noise ratio (SNR) on the short
timescales on which the visibility phase varies.  Self-calibration
enables corrections of unmodeled phase errors, but destroys absolute
positional information.  As such, astrometric observations must use
phase-referencing, where scans are alternated on the (possibly weak)
target and a strong nearby calibrator (the ``nodding calibrator'').
Phase connection across scans on the nodding calibrator allows the
visibility phase to be interpolated between the scans on the target,
extending the effective coherence time from minutes to hours. This
allows astrometry of weak sources, relative to the (assumed
fixed) position of the nodding calibrator. This scheme requires limits
on both the angular separation between the target and the nodding
calibrator, and the cycle time between them.  These requiremnts are
outlined, for example, by
\citet{VLBA_24},
who recommend cycle times ranging from 30~seconds at 43~GHz to
300~seconds at 1.4~GHz, with an angular separation $\lesssim
5\arcdeg$.

The negative spectral index of pulsars can be exploited by
observations at lower frequencies, where they are stronger. However,
the achieved resolution is proportional to the synthesized beam size,
and inversely proportional to the SNR. At lower frequencies, the beam
is larger, and the system temperatures are generally higher, offseting
some of the gain in pulsar signal. Additionally, ionospheric effects
become more troublesome at lower frequencies. Our chosen compromise
was to observe at L-band (1.4--1.7~GHz), with a typical
target--calibrator separation of 5\arcdeg\ and a cycle time of 5
minutes. Even with this strategy, the residual ionospheric errors
after phase-referencing can occasionally be large enough to prevent
correct phase interpolation and useful astrometry.

Various techniques have been explored to calibrate out differential
ionospheric effects, including the use of GPS data
\citep[e.g.][]{RMG+00}.  For PSR B0919+06, we use an in-beam
calibrator, as discussed in detail in \citet{FGBC99}: there is a faint
($\sim 10$ mJy) source, J0922+0638, 12\arcmin\ from the pulsar,
within the $\sim$30\arcmin\ primary beam of a VLBA antenna at
1.4~GHz. The pulsar data are correlated twice, once at the pulsar
position and once at the position of the in-beam calibrator. We
phase-reference the data to the nodding calibrator, self-calibrate the
visibility phases so that they are consistent with the visibility
function expected for the in-beam calibrator, and transfer these
corrections to the pulsar data to carry out the imaging and astrometry.

This procedure has two crucial advantages over ordinary
phase-referencing. Phase errors from the different ionospheric
electron column densities along different lines of sight generally
increase with angular separation. Using an in-beam calibrator not only
reduces the target--calibrator angular separation from $\sim$5\arcdeg\
to a few arcminutes, but the visibility phase corrections are also
derived for the actual times when the pulsar was observed, and thus
avoid interpolation over time. This procedure reduces the residual
ionospheric errors so that sub-\mas\ astrometry is achievable.

\subsection{Imaging and Astrometry}

The VLBA data used here were obtained as part of two different
programs, using different nodding calibrators and slightly different
frequency setups. The earlier observations (1994--1996) used VLA
calibrator 0906+015 (6\arcdeg\ away from the target) as a nodding
calibrator, and a cycle time of 4 minutes on target and 2 minutes on
the calibrator. The later observations (1998--2000) used J0914+0245
(4\arcdeg\ away) as the calibrator and shortened the cycle time to 3
minutes on target, 2 minutes on calibrator. We also adopted a strategy
of snapshot images at two closely spaced epochs (3 to 7 days) in order
to verify the consistency of the astrometry, and slightly adjusted the
specific observation frequencies (8 independent channels of 8 MHz
each) within the 1.4--1.7~GHz band to avoid some known radio frequency
interference. During some later epochs, the pulsar was gated at the
VLBA correlator, boosting the SNR by $\sim f^{-1/2}$, typically a
factor of 3--4, where $f = T_{\rm on}/(T_{\rm on} + T_{\rm off})$ is
the gate duty cycle.

We note that the changes in observing setup between epochs do not
affect the astrometry, since the final pulsar positions are determined
relative to the (fixed) position of the in-beam calibrator, which was
assumed to be the same at all epochs. The absolute position of the
in-beam calibrator, determined with respect to J0914+0245, is known
with a precision of $\sim$10 mas.

\begin{figure}[ht]
\epsscale{0.6}
\plotone{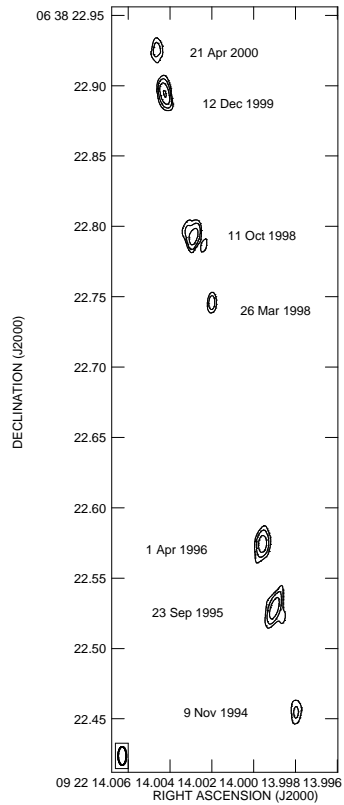}
\caption{Proper motion of PSR B0919+06, depicted as images for a
series of epochs. Contours are shown at $5, 10, 20, 40 \sigma$, where
$\sigma$ is the RMS map noise. There are no pixels in the map at or
below $-5 \sigma$.  The period-averaged pulsar flux density varied
between 2 and 15~mJy while the RMS noise ranged from 0.1 to
0.9~mJy.}
\label{Fig:psrpm}
\end{figure}

The data were reduced using standard VLB phase-referencing procedures
\citep{BC95} using {\tt AIPS}, the Astronomical Image Processing
System.  This involved amplitude calibration using the system
temperature at each antenna after flagging bad records; {\em a priori}
phase calibration using pulse calibration tones at the VLBA when
available; fringe-fitting to the nodding calibrator; and
self-calibration of the in-beam calibrator. The calibration was
transferred to the pulsar, and both the in-beam source and the pulsar
were imaged.  During epochs with a relatively undisturbed ionosphere,
the in-beam source was observed to be point-like before
self-calibration. Self-calibration yields a point source image for the
in-beam source, whose position is used for relative astrometry of the
pulsar.  Figure~\ref{Fig:psrpm} shows one pulsar image from each of
the seven different epochs, mosaiced together to illustrate the proper
motion. 

The {\tt AIPS} task {\tt JMFIT} was used to fit elliptical Gaussians
to the images in order to obtain astrometric positions and position
errors, as listed in Table~\ref{Table:offsets}.  The position errors
listed for each epoch have three components, which we estimate and add
in quadrature: (a) random errors in the position of the in-beam
calibrator, estimated as the beam full width at half-maximum
(FWHM)$/(2\times\mbox{SNR})$, (b) random errors in the position of the
pulsar, estimated as the fitted Gaussian FWHM$/(2\times\mbox{SNR})$,
and (c) systematic errors in the pulsar position, contributed by the
residual ionosphere between the in-beam calibrator and the target, as
well as other possibly unknown errors.  The deviation of the observed
pulsar image from that of a point source convolved with the dirty beam
(or point spread function) provides a measure of the residual
systematic errors.  We take the deconvolved size of the pulsar
($\theta_d$, which is zero when the image is truly point-like) and use
the quantity $\theta_d/\sqrt{(N_{\rm ant}-1) T_{\rm obs}/T_{\rm
iono}}$ as an estimate of this last error component. This quantity is
not sensitive to the choice of $T_{\rm iono}$, the ionospheric
coherence time, which is empirically determined as $\sim 0.15$
hour. The number of antennas $N_{\rm ant}$ is usually 10 (or 9 when we
lost a VLBA station due to inclement weather) while the observation
time $T_{\rm obs}$ ranged from 1 to 3 hours.  The small separation
(12\arcmin) of the pulsar and the in-beam calibrator implies that any
systematic errors in the correlator model \citep{MAE+98} are
negligible, and any residual systematic errors in the astrometry are
likely to be small ($<0.1$ mas).

\begin{deluxetable}{ccc}
\tablecolumns{3}
\tablewidth{0pc} 
\tablecaption{Astrometry Measurements for PSR B0919+06\label{Table:offsets}}
\tablehead{ 
\colhead{}    &  \multicolumn{2}{c}{Pulsar Position
Offset\tablenotemark{a}} \\ \cline{2-3} 
\colhead{Observation Date} & \colhead{East--West (mas)}
&\colhead{North--South (mas)} 
}
\startdata

1994.857 $\ldots$ & -45.89 $\pm$ 0.64 & -263.20 $\pm$ 1.41 \\
1995.728 $\ldots$ & -30.69 $\pm$ 0.33 & -189.60 $\pm$ 0.59 \\
1996.252 $\ldots$ & -22.20 $\pm$ 0.38 & -143.80 $\pm$ 0.76 \\
1998.233 $\ldots$ & 13.86 $\pm$ 0.31 & 27.70 $\pm$ 0.73 \\
1998.244 $\ldots$ & 14.60 $\pm$ 0.32 & 28.30 $\pm$ 0.77 \\
1998.778 $\ldots$ & 26.52 $\pm$ 0.89 & 75.40 $\pm$ 1.02 \\
1998.797 $\ldots$ & 27.27 $\pm$ 0.97 & 77.20 $\pm$ 1.35 \\
1999.917 $\ldots$ & 46.34 $\pm$ 0.20 & 173.20 $\pm$ 0.52 \\
1999.947 $\ldots$ & 46.93 $\pm$ 0.17 & 175.80 $\pm$ 0.38 \\
2000.307 $\ldots$ & 52.30 $\pm$ 0.49 & 207.30 $\pm$ 1.03 \\
\enddata 

\tablenotetext{a}{Measured from 
$09^{\rm h}22^{\rm m}14\fs0011$, $+06\arcdeg38\arcmin22\farcs7180$, 
referenced to in-beam position 
$09^{\rm h}23^{\rm m}03\fs8991$, $+06\arcdeg38\arcmin58\farcs9980$ 
(J2000.0).} 

\end{deluxetable} 

\subsection{Likelihood Analysis for Astrometric Parameters\label{Subsec:lik}}

To derive the best fit values and confidence intervals for the
parallax and proper motion of PSR B0919+06, we calculate a likelihood
function for the parameters that describe the apparent pulsar position
as a function of time. These parameters include ($\Delta\alpha_0,
\Delta\delta_0$), the offset of the pulsar from an arbitrary assumed 
position at the reference epoch (1997.0); ($\mu_{\alpha},
\mu_{\delta}$), the proper motions in right ascension and declination;
and $\pi$, the annual trigonometric parallax.

The likelihood function is calculated as follows:
\beq
{\cal L} = \prod_{i=1}^{N_{\rm data}} g\left((\alpha_i -
\hat{\alpha_i}) / \sigma_{\alpha_i}\right) g\left((\delta_i -
\hat{\delta_i}) / \sigma_{\delta_i}\right), 
\eeq
where
\begin{displaymath}
\hat{\alpha_i}, \hat{\delta_i} = \hat{\alpha_i}, \hat{\delta_i} 
\left( \Delta\alpha_0, \Delta\delta_0, \mu_{\alpha}, \mu_{\delta},
\pi;\; t_i \right)   
\end{displaymath}
are model estimates and $\alpha_i, \delta_i$ are the observed values
of the pulsar coordinates at epoch $t_i$ (offsets from the 
assumed position for 1997.0, as listed in Table~\ref{Table:offsets}),
$N_{\rm data}$ is the number of data points (10 in this case), 
and $g$ is a normalized Gaussian function with zero mean and unit
variance.

To obtain the marginal probability distribution of each parameter
$\vartheta \in \thetavec = (\Delta\alpha_0, \Delta\delta_0,
\mu_{\alpha}, \mu_{\delta}, \pi)$, we calculate the normalized
integral of the likelihood function over all other parameters:
\beq
f_{\vartheta}(\vartheta) = \frac{ \int_{\mbox{\tiny exclude\ }
\vartheta} d\thetavec \, {\cal L}(\thetavec)} {\int
d\thetavec \, {\cal L}(\thetavec)}.
\eeq
From the marginal distributions for each parameter, we obtain the 
best fit (median) values, as well as 68\% ($1\sigma$) confidence
intervals. These are summarized in Table~\ref{Table:results}, along
with derived estimates for the distance $D$ and transverse velocity
$\Vpperp$. 

\begin{deluxetable}{lcc}
\tablecolumns{3}
\tablewidth{0pc} 
\tablecaption{Astrometric Parameters for PSR B0919+06\label{Table:results}}
\tablehead{ 
\colhead{Parameter} & \colhead{Median value} & \colhead{$1\sigma$ Error}
}
\startdata
$\mu_{\alpha}$ (mas yr$^{-1}$) & 18.35 & 0.06 \\
$\mu_{\delta}$ (mas yr$^{-1}$) & 86.56 & 0.12 \\
$\pi $ (mas)            & 0.83  & 0.13 \\
\\ 
D (kpc)                 & 1.21  & 0.19 \\
$\Vpperp$ (\kms)        & 505   & 80 \\
\enddata 
\end{deluxetable} 

\begin{figure}[ht]
\epsscale{0.9}
\plotone{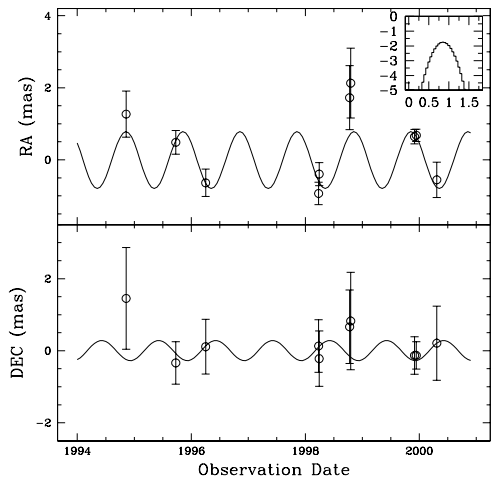}
\caption{Residual position offsets in right ascension and declination
after subtracting the proper motion; sinusoids corresponding to the
best fit parallax of 0.83~mas are overplotted. Inset: The marginal
probability distribution for the parallax, $\log_{10} f_{\pi}(\pi)$
plotted against $\pi$ (mas).  A null result is excluded by $\sim$8 orders of
magnitude, or over 6$\sigma$. }
\label{Fig:psrparallax}
\end{figure}

Figure~\ref{Fig:psrparallax} shows the residual position offsets after
subtracting the best fit proper motion, with the best fit parallax
sinusoids overplotted. The marginal probability distribution of the
parallax is plotted as an inset in Figure~\ref{Fig:psrparallax}.  The
parallax, $\pi=0.83 \pm 0.13$~mas, differs significantly from
\citet{FGBC99}, who report $\pi=0.31 \pm 0.14$~mas using a subset of
the current data set (4 epochs from 1994--1998). It is now apparent
that the position errors were underestimated, leading to a parallax
estimate $4\sigma$ displaced from the current result.  To test the
robustness of the fit reported here, the likelihood analysis was
repeated after eliminating data points singly (1994, 2000) and in
pairs (1994 and 2000, both points from October 1998). In all cases,
the best fit parameters remained consistent (though with greater
uncertainty) with the values for the fit including all data. For
example, the 68\% confidence interval for the parallax ranged from
$0.81 \pm 0.15$~mas to $0.87 \pm 0.15$~mas, compared to $0.83 \pm
0.13$~mas when using all the data.

We also note that the likelihood of zero parallax is $\sim$8 orders of
magnitude lower than the likelihood of $\pi=0.83$~mas, and the null
result is ruled out at over the $6\sigma$ level. This result implies
that even if B0919+06 were twice as far away ($\sim 2.4$~kpc), the
parallax signature of 0.4~mas could have been detected at $3\sigma$.
With a suitable in-beam calibrator and the addition of more sensitive
antennas (Arecibo, GBT) to the VLBA, a parallax signature should be
detectable to 10~kpc at $5 \sigma$ with $\sim 25$ observations. Beyond
this, the variable source structure of calibrators at the 0.1~mas
level \citep{FCF96} places a basic limit on current astrometric
techniques.

\subsection{Transverse Velocity of PSR B0919+06}

From our measurement of the total proper motion ($88.48 \pm 0.13$
mas yr$^{-1}$) and parallax ($0.83 \pm 0.13$ mas) of B0919+06, we can
estimate its transverse velocity to be $505 \pm 80$ \kms. This is
consistent with the mean pulsar population velocity of $\sim 450$
\kms\ estimated by \citet{LL94}, as well as the two-component model
with characteristic speeds of 175 and 700 \kms\ obtained by
\citet{CC98}.


\section{Interstellar Scintillation: Data and Analysis\label{Sec:ISS}}

Electron density fluctuations along the line of sight through the
interstellar matter (ISM) scatter radio signals. For pulsars, this
phenomenon has been studied extensively through dynamic spectra
(intensity variations as a function of time and frequency), whose
characteristics allow inferences about the distribution of scattering
material along the line of sight. Much of the information in dynamic
spectra can be condensed into two parameters derived from the
autocorrelation function (ACF): the scintillation timescale $\dtd$
(defined as the $1/e$ width of the ACF intercept on the time lag
axis) and the decorrelation bandwidth $\dnud$ (the half-width at half
maximum of the ACF intercept on the frequency lag axis).

\begin{figure}[ht]
\epsscale{0.9}
\plotone{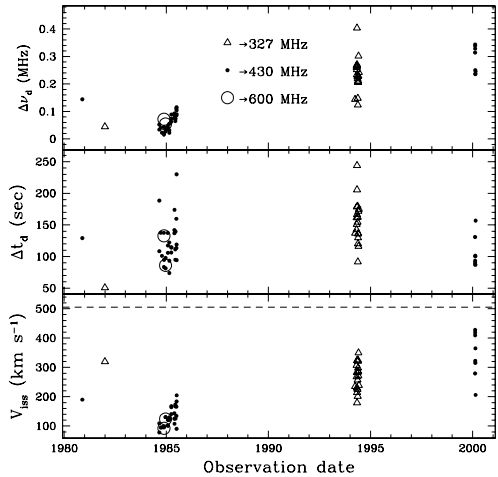}
\caption{Multifrequency scattering parameters for PSR B0919+06: (Top)
decorrelation bandwidths $\dnud$; (Middle) scintillation timescales
$\dtd$; and (Bottom) scintillation speed $\Vissu$ obtained over 20
years (see text for references). The pulsar speed derived from VLBI
parallax and proper motion is plotted with a dashed line.}
\label{Fig:scattpars}
\end{figure}

For PSR B0919+06, measurements of these parameters exist over a 20
year timespan, from 1980 to 2000 \citep[and unpublished data from
Arecibo Observatory in 2000]{CWB85,CW86,BRG99}.  \citet{CW86} observed
this pulsar in 1984--85, and report interference fringes due to
multipath propagation effects in some of their dynamic spectra,
signifying that the ISM had caused multiple imaging of the pulsar,
while other epochs show random structure consistent with single-image
diffractive scattering.  \citet{BRG99} observed this pulsar during
1994, and report gradual and systematic variations in the dynamic
spectra, but no fringing events.  The derived scintillation parameters
for the entire set of data are plotted in Figure~\ref{Fig:scattpars}
(top and middle panels): as expected, these parameters vary
significantly with time.

In the analysis that follows, we do not address either the statistical
errors in quantifying $\dnud, \dtd$ due to the finite number of
scintles obtained in each observation or the measurement
uncertainties due to the finite number of independent experiments.
These errors are adequately represented by the scatter in the data,
since we treat each observation independently. Refractive effects may
also modify $\dnud$ and $\dtd$, particularly in the epochs where
multipath effects are important. These effects are difficult to
quantify, but we account for their presence in the interpretation.

\subsection{Scintillation and Scattering Formalism}

For the analysis, we adopt the theoretical framework of
\citet[hereafter CR98]{CR98}, who provide a general treatment of
diffractive ISS without assuming any specific scattering geometry
(like a thin screen or uniform medium). This generality neccessarily
leads to some algebraic complexity, which proves worthwhile in the
analysis.

The scintillation speed $\Viss$ is defined as the speed of the ISS
diffraction pattern with respect to the observer, which includes
contributions from the velocities of the earth and the medium as well
as the pulsar. Given values for $D$ (in kpc) and observing frequency
$\nu$ (in GHz), and measurements of $\dnud$ (MHz) and $\dtd$ (sec),
we obtain $\Viss$ in \kms\ (CR98, eq. 13): 
\beq
\Vissu = \Aissu \frac{\sqrt{D \dnud}}{\nu \dtd},
\eeq
where $\Aissu = 2.53 \times 10^{4}$ \kms, and the subscripts 5/3 and u
denote a Kolmogorov scattering medium that is statistically
uniform. The estimated scintillation speed is plotted for the 
data set in the bottom panel of Figure~\ref{Fig:scattpars}.

To estimate the transverse speed of the pulsar $\Vpperp$ from $\Viss$,
we assume that the speeds of the observer and the medium are small
compared to the pulsar's; then (CR98, eq. 23): \beq \Vpperp =
\Wc\Wdpm\Vissu, \eeq where $\Wc$ and $\Wdpm$ are weighting factors
which relate the scintillation pattern speed to the pulsar speed, and
rescale the velocity from the uniform Kolmogorov medium case to the
actual medium. $\Wc$ depends on the wavenumber spectrum of the medium
as well as its distribution (CR98, eq. 18), and its contribution
is small (maximum range of 0.9--1.3) unless one assumes solely a thin
screen for the distribution of the entire scattering medium.  For the
remaining analysis, we assume this quantity is unity.

$\Wdpm$ is a weighting factor determined by the distribution of the
scattering material along the line of sight, which is proportional to
the coefficient of the electron-density wavenumber spectrum
$\Cnsqr(s)$ (CR98, Appendix A), where $s$ is the distance measured from
the pulsar towards the observer, and varies from 0 to $D$. An exact
expression for $\Wdpm$ is derived for a square law phase structure
function in CR98 (eq. 25):
\beq
\Wdpm(D) \equiv 
\left[\frac
            {2\int_0^D ds\, (s/D) (1-s/D) \Cnsqr(s)}
            {\int_0^D ds\, (1-s/D)^2 \Cnsqr(s)}
        \right]^{1/2}.
\label{Eqn:wdpm}
\eeq
This expression is only approximately valid for a true Kolmogorov
medium, but the approximation is sufficient for our analysis.

\subsection{Hybrid Analysis of VLB and ISS Data}

There are two independent measurements of the transverse velocity of the
pulsar, one from the observable VLB proper motion $\mu$, and the other
from the measured scintillation parameters $\dnud, \dtd$ at an
observation frequency $\nu$:
\begin{eqnarray}
\Vpperp & = & \mu D 
\label{Eqn:Vmu} \\
\Vpperp & = & \Wc\Wdpm\Aissu \frac{\sqrt{D \dnud}}{\nu \dtd}.
\label{Eqn:Viss}
\end{eqnarray}
$\Wdpm$ depends on the (unknown) distribution of scattering material,
as expressed in equation (\ref{Eqn:wdpm}), while $D$ is usually known
only roughly from dispersion measure--distance models (TC93).

In general, if a parallax distance is unavailable for a pulsar, 
the different dependencies of $\Vpperp$ on $D$ can be used to
iterate the two equations (\ref{Eqn:Vmu},\ref{Eqn:Viss}) for a
given distribution of scattering material (uniform, exponential,
TC93), and thus solve for the distance as well as the transverse
velocity. This procedure is especially useful for pulsars well above
the Galactic $z$ scale height, where dispersion measure-based models
can provide only lower limits on the distance, and parallaxes are not
likely to be obtained in the forseeable future (as discussed in
\S\ref{Subsec:lik}). 

Alternatively, it is possible to refine electron density models by
postulating the existence of extra scattering material along the line
of sight, in the form of a clump or a screen which contributes
increments in dispersion measure $\dDM$ and scattering measure $\dSM$
at a location $D_s$. Then a set of constraints on the acceptable
values of the screen parameters ($\dSM, D_s$) as well as the (poorly
known) distance $D$ can be derived from this hybrid analysis.

\subsection{Hybrid Analysis using the Parallax Distance for B0919+06}

For PSR B0919+06, the measured parallax distance allows stronger
constraints on the scattering geometry.  The observation of fringing
events for this pulsar, as well as the wide variability in its
scintillation parameters \citep{CW86,BRG99}, suggest that this pulsar
is viewed through a localized screen or clump of variable scattering
strength in addition to a distributed medium. Using only the TC93
electron density model results in an overestimate of the distance to
this pulsar, and fails to account for either the observed values of
the scattering parameters or their variation with time
(Figure~\ref{Fig:scattpars}).

We note that large changes in scintillation parameters can be produced
with relatively small changes in the average electron density at each
epoch. From the definitions of dispersion measure and scattering
measure (DM $= \int_0^D ds\, n_e(s)$; SM $= \int_0^D ds\, \Cnsqr(s)$),
we have:
\begin{eqnarray} 
\dDM &=& n_{e,s} \Ds ,\label{Eqn:dDM}\\ 
\dSM &=& \Cnsqrs\Ds ,\\
\Cnsqrs &\propto& F_s n_{e,s}^2, \label{Eqn:Cnsqr}
\end{eqnarray} 
where $F$ is a dimensionless ``fluctuation parameter'' as defined in
\citet{TC93}, $n_e$ is the electron density in the medium, the
subscript $s$ designates the values for a screen of thickness $\Ds$,
and $\dDM, \dSM$ are epoch-dependent values.  In order to fit the
observations, we postulate the existence of a thin patchy screen of
scattering material along the line of sight in addition to the TC93
scattering material:
\beq
\Cnsqr(s) = \Cnsqrtc(s) + \dSM \, \delta (s-D_s),
\eeq
where the thin screen contributes an additional scattering measure
$\dSM = \Cnsqrs\Ds$ at a location $s=D_s$, represented by the delta
function $\delta (s-D_s)$. Recasting equations
(\ref{Eqn:dDM}--\ref{Eqn:Cnsqr}) in terms of the
differential DM contributed by the screen, and putting in the
appropriate scaling factors, 
\beq 
\dSM = \left[(1/3)(2\pi)^{-1/3}\right] K_u F_s 
\frac{(\dDM)^2}{\Ds}, \label{Eqn:dSM_dDM}
\eeq
where the scale factor $K_u = 10.2 \times 10^{-3} \mbox{m}^{-20/3}
\mbox{cm}^6$ provides the appropriate unit conversion for $\dSM$ in
kpc~m$^{-20/3}$, $\dDM$ in pc~cm$^{-3}$ and $\Ds$ in parsecs.

We can rewrite the weighting function $\Wdpm$ from equation
(\ref{Eqn:wdpm}) in terms of the TC93 model and this additional
scattering screen: 
{\small
\begin{eqnarray}
\lefteqn {\Wdpm(D) \equiv} \label{Eqn:wtc_plus_screen}  \\ 
 & & \hspace{-25pt} \left[\frac
{2\int_0^D ds\, (s/D)(1-s/D) \Cnsqrtc(s) +  (D_s/D)(1-D_s/D) \dSM}
{\int_0^D ds\, (1-s/D)^2 \Cnsqrtc(s) + (1-D_s/D)^2 \dSM}
\right]^{1/2}. \nonumber
\end{eqnarray}
}
We integrate the TC93 model out to the parallax distance derived
for this pulsar, which gives $\Wdpm$ as a function of two screen
parameters in equation (\ref{Eqn:wtc_plus_screen}), i.e.~the relative 
scattering strength $\dSM/\mbox{SM}$, and location $D_s/D$.

Additionally, from equations (\ref{Eqn:Vmu},\ref{Eqn:Viss}), we solve
for the values of $\Wdpm$ required to match the scintillation and
interferometric velocity estimates (with $\Wc$ set to unity):
\beq
\Wdpm = \frac{\mu \nu \dtd}{\Aissu} \left(\frac{D}{\dnud}\right)^{1/2}.
\label{Eqn:w_observed}
\eeq

\begin{figure}[ht]
\epsscale{0.9}
\plotone{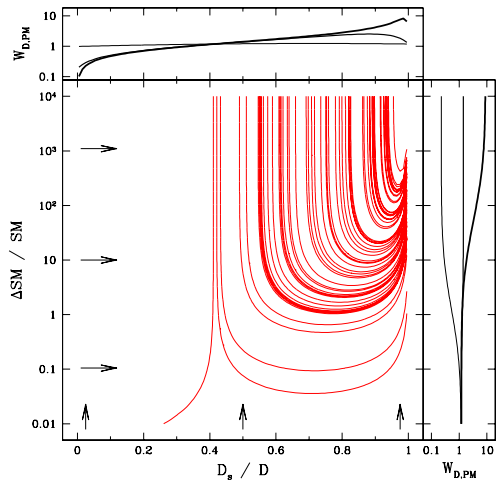}
\caption{$\Wdpm$ expressed as a function of scattering screen strength
$\dSM$ and distance $D_s$ from the pulsar towards the observer. The
contours represent the value of $\Wdpm$ required at each epoch shown
in Figure~\ref{Fig:scattpars} for the scintillation speed to match the
actual $\Vpperp$. Panels show slices through the surface on which
contours are drawn, taken at the locations of the arrows. Top panel:
$\dSM/\mbox{SM} = 0.1, 10, 1000$ (thin to thick line); side panel:
$D_s/D = 0.025, 0.050, 0.975$ (thin to thick line). Assuming a single
screen, its location $D_s/D$ must be consistent with all of the
contours (barring strong refractive scintillation effects), while the
range of contours along the $\dSM/\mbox{SM}$ axis gives the required
variation in the scattering strength of the screen.}
\label{Fig:screen}
\end{figure}

Thus $\Wdpm$ is known as a function of $\dSM/\mbox{SM}$, $D_s/D$
(eq. \ref{Eqn:wtc_plus_screen}), and the values of $\Wdpm$ required
for the two estimates of $\Vpperp$ (eqs. \ref{Eqn:Vmu},\ref{Eqn:Viss})
to agree at each observation are specified by equation
(\ref{Eqn:w_observed}).  This information is summarized in
Figure~\ref{Fig:screen}, where the required values of $\Wdpm$ are
plotted as contours against $\dSM/\mbox{SM}$ and $D_s/D$ on the
surface defined by equation (\ref{Eqn:wtc_plus_screen}). The range of
values required for the scattering strength and location of the screen
(in addition to the TC93 model) can thus be determined.

\subsection{Results from the Hybrid Analysis} 

A reasonable assumption is that a turbulent screen may fluctuate in
scattering strength on short timescales, while its location varies
slowly (if at all). In that case, the screen must be located at a
distance consistent with {\em all} of the observed contours in
Figure~\ref{Fig:screen}, while the spread in the contours along the
$\dSM/\mbox{SM}$ axis gives a measure of the range over which the
screen scattering strength has varied over the 20 years spanned by
observations. From the figure, consistency with the innermost contours
requires $D_s/D \gtrsim 0.9$. However, the multiple imaging and
fringing events observed during 1986 indicate the occurrence of strong
refractive scintillation effects. The current analysis of {\em
diffractive} scintillation probably overestimates the scattering
required during that epoch. Thus the innermost set of contours in
Figure~\ref{Fig:screen} may not be useful in locating the screen, and
the constraints on the screen can be relaxed to $D_s/D \gtrsim 0.8$,
with $\dSM/\mbox{SM}$ ranging from 0.1 to 100.

For a pulsar distance of 1.2 kpc, this result suggests that the
scattering screen is within $\sim$240 parsecs of the Earth,
implicating structure within the local ISM.  Assuming a screen
thickness $\Ds=10$ parsecs, equation (\ref{Eqn:dSM_dDM}) gives an
approximate range for $\dDM$ and the fluctuation parameter $F_s$: \beq
10^{-5} \lesssim F_s(\dDM)^2 \lesssim 10^{-2} (\mbox{pc cm}^{-3})^2.
\eeq Assuming, for example, that $F_s=10$ (consistent with values in
TC93, Table 2), the required range in $\dDM$ is $10^{-3}$ to $3 \times
10^{-2}$ pc cm$^{-3}$.  For the assumed screen thickness of 10
parsecs, this requires $n_{e,s}$ between $10^{-4}$ and 0.003
cm$^{-3}$. These values are consistent with the DM variability
observed for several nearby pulsars ($\dDM_{RMS} \sim \mbox{few}\times
10^{-3}$ pc cm$^{-3}$), including B0919+06 itself, which had a DM
variation of $4\times 10^{-3}$ pc cm$^{-3}$ in 1989--1991
\citep{PW92}.  Of course, in reality, both $F_s$ and the screen
thickness $\Ds$ may vary along with changes in electron density, but
these values show that the derived constraints on screen location and
$\dSM/\mbox{SM}$ are well within the range of physical
possibilities. Additionally, simultaneous measurements of DM
variability and changes in ISS parameters can be used to estimate
$\Ds$, which is related to the outer scale size of ISM fluctuations.

\subsection{Interpreting the Screen in terms of the Local ISM}

Combining the analysis for PSR B0919+06 with parallaxes and scattering
measurements for other pulsars in the third Galactic quadrant produces
a coherent picture of the local ISM in this direction.  Measurements
of Galactic coordinates ($\ell, b$), distance, and DM for
third-quadrant pulsars are summarized in Table~\ref{Table:pulsars},
along with derived values of $n_e$, angular separation $\Delta\theta$
and Galactic $z$ height.  \citet{TBM+99} review the local ISM, and
their Figure~2 provides useful background for our discussion. The
preceding analysis, and a recent (upward) revision in the distance to
PSR~B0950$+$08 \citep{BBB+00}, allows us to improve upon their
discussion for the third Galactic quadrant.

It has long been known that there is an elongated cavity in the local
neutral ISM, surrounding the Sun and extending several hundred parsecs
towards Galactic longitude $\ell=240\arcdeg$. This feature appears to be
correlated with the region of low reddening in the Gould Belt, between
$\ell=210\arcdeg$ and $255\arcdeg$ \citep{L78}, and 
the much-reduced \HI absorption at both high and low latitudes in this
longitude range \citep{P84}. The Local Bubble fills part of this
cavity out to a distance of~0.1--0.2~kpc \citep[e.g.,][]{CR87}, though
the Bubble boundary is not well defined in this direction. Beyond the
Local Bubble, \citet{H98} has proposed a superbubble, GSH~238+00+09,
along $\ell=238\arcdeg$. This extends from~0.2 to~1.3~kpc from the
Sun, and fills another part of the elongated cavity.

It is expected that the region between these two structures has clumps
of dense and partially ionized gas, possibly left over from the
creation of either (or both) of the bubbles, or from their interaction
with each other and their environments.  For example, \citet{DG98}
report the existence of two dense clouds of \HII between 40 and 90 pc
from the sun, along the line of sight towards $\beta$ Canis Majoris
($\ell,b = 226\arcdeg, -14\arcdeg$), and speculate that these
originate in the formation of the Local Bubble.  \citet{H98}
postulates ``filaments'' of warm ionized material on the surface of
superbubbles.  The scattering properties of various pulsars toward the
third Galactic quadrant (Table~\ref{Table:pulsars}) are consistent
with the existence of such a turbulent interface region.

The most striking feature of Table~\ref{Table:pulsars} is the
difference in $n_e$ between the lines of sight to PSRs~B0950$+$08 and
B0919$+$06.  Although separated by only 7\fdg8 (35~pc at a distance of
250~pc), PSR~B0919$+$06 exhibits strong scattering and has undergone
episodes of multiple imaging and refractive fringing, while
PSR~B0950$+$08 has one of the lowest scattering measures known
\citep{PC92}.  There are two possible causes for the difference in the
properties of these two pulsars.

\begin{deluxetable}{lcccc}
\tablecolumns{5}
\tablewidth{0pc} 
\tablecaption{Pulsars in the Third Galactic Quadrant\label{Table:pulsars}}
\tablehead {
 \colhead{PSR} & \colhead{B0919$+$06} & \colhead{B0950$+$08} & 
 \colhead{J1024$-$0719} & \colhead{B0823$+$26}
}
\startdata
$\ell$ (\arcdeg)  & 225.42   & 228.91   & 251.70   & 196.96 \\
$b$ (\arcdeg)    & $+36.39$ & $+43.70$ & $+40.52$ & $+31.74$ \\
\\
DM (pc~cm${}^{-3}$) & 27.31    & 2.97         & 6.50            & 19.48    \\
$D$ (kpc)        & 1.2\tnm{1} & 0.28\tnm{2}  & $< 0.23$\tnm{3} & 0.38\tnm{4}\\
$z$ (kpc)        & 0.72        & 0.19         & $< 0.15$        & 0.20  \\
$n_e$ (cm${}^{-3}$) & 0.023    & 0.011        & $> 0.029$       & 0.055  \\
\\
$\Delta\theta$ (\arcdeg) & 0 & 7.8 & 21.4 & 23.7 \\
$\Delta l_{\perp} = D_s \Delta\theta$ (pc) & 0 & 35 & 93 & 104 \\
\enddata
%
%
\tablerefs{Distances from (1) this work; (2) Brisken \etal\ 2000; 
(3) Toscano \etal\ 1999b (upper limit derived from 
$\mu$, P, and $\dot{\mbox{P}}$); and (4) Gwinn \etal\ 1986. }
\end{deluxetable}

One possibility is that PSR~B0950$+$08 may be closer than the
turbulent interface region.  This model requires the distance to the
interface region to be larger than $280 \pm 25$~pc, the (revised)
distance to PSR~B0950+08, or equivalently, $D_s/D < 0.8$. We concluded
above that $D_s/D \gtrsim 0.8$, but given the uncertainties in
incorporating the effects of strong \emph{refractive} scintillation
into the estimate for $D_s/D$, we can not rule out this
possibility. Rather, we take $D_s/D \simeq 0.75$--0.8 as a reasonable
range for the screen distance.

An alternative possibility is that PSR~B0950+08 is within or beyond
the turbulent interface region, but the turbulence is concentrated
into clumps.  In order to cover PSR~B0919+06 but not PSR~B0950+08, the
typical (transverse) scale size of a clump would have to be less than
roughly 30~pc.  This suggestion is consistent with the 10 pc (radial)
scale size of the scattering screen that was assumed in our analysis.

The lines of sight to pulsars B0823+26 and J1024$-$0719 also provide
useful information, though the derived constraints are weaker.  Both
pulsars have mean electron densities along the line of sight
comparable to or larger than B0919+06.  Both exhibit stronger
scattering than B0950+08, particularly B0823+26 which has undergone
episodes of strong refractive fringing and multiple imaging
\citep[e.g.,][]{CFC93}.  J1024$-$0719 is closer to the Sun than
B0950+08, perhaps indicating that B0950+08 is within or beyond the
turbulent interface region.  However, both B0823+26 and J1024$-$0719
are near the edges of the \HI cavity.  Whether their scattering
properties and mean electron densities are due to the ``edge'' of the
\HI cavity or to the interface between the Local Bubble and the
GSH~238+00+09 superbubble is not clear.

We conclude that interstellar scintillometry of PSR~B0950+08 and
PSR~B0919+06 indicates a turbulent interface region between the Local
Bubble and the GSH~238+00+09 superbubble.  This turbulent interface
region begins 250--300~pc from the Sun in the direction $(\ell, b)
\approx (225, 40)$ and possibly contains clumps of scale size $<
30$~pc. This result is not inconsistent with the model of
\citet{BGR98}.  Given the height of PSR~B0950+08 above the Galactic
plane ($\sim 200$ pc), it is possible that the interface region, and
therefore the Local Bubble and the GSH~238+00+09 superbubble, extend
to higher Galactic latitudes.

\section{Conclusions}

Parallax distances to pulsars have the prospect of resolving several
outstanding questions, both about the pulsar population and about the
intervening ISM. In this work, we have presented VLBA astrometry on
PSR B0919+06, using phase-referenced observations with an in-beam
calibrator only 12\arcmin\ from the pulsar. The use of the in-beam
source reduces ionospheric effects and other astrometric errors to the
0.1 mas level and allows sub-\mas\ astrometry.  With ten observations
(seven distinct epochs) distributed over seven years, we derive a
proper motion $\mu_{\alpha} = 18.35 \pm 0.06$ mas yr$^{-1}$,
$\mu_{\delta} = 86.56 \pm 0.12$ mas yr$^{-1}$, and measure a parallax
$\pi = 0.83 \pm 0.13$ mas (68\% confidence intervals). This result
implies a transverse speed of $505 \pm 80$ \kms\ at a distance $D =
1.21 \pm 0.19$ kpc, making PSR B0919+06 one of the farthest objects
for which a trigonometric parallax has been obtained. The use of an
in-beam calibrator with the VLBA at 1.5~GHz can provide trigonometric
parallaxes to twice this distance, or even further (to 10~kpc)
with the use of more sensitive antennas and higher frequencies.

For the measured distance of 1.2~kpc and a DM of 27.31 pc~cm$^{-3}$
\citep{PW92}, the derived mean electron density towards B0919+06 is
0.023 cm$^{-3}$. This line of sight is along the same Galactic
longitude as the superbubble proposed by \citet{H98}, as well as
clouds of ionized gas in the local neighbourhood \citep{DG98}, though
at a higher Galactic latitude. We use scintillation data spanning 20
years in a hybrid analysis, equating the derived ISS and
interferometric speeds of the pulsar in order to constrain the
distribution of scattering material along the LOS. We find that excess
scattering material is required compared to the standard TC93 model,
and assuming that this excess material is confined to a thin screen or
clump, we constrain its location to within $\sim 250$~pc of the Sun.
Comparison with the neighboring lines of sight to pulsars B0950+08,
J1024$-$0719, and B0823+06 indicates that this result is consistent
with a turbulent interface region between the Local Bubble and the
GSH~238+00+09 superbubble, which possibly contains clumps of scale
size $< 30$~pc.  We note that using an updated model of the Galactic
electron density distribution (Cordes {\&} Lazio, in preparation) will
not materially alter these conclusions.

\acknowledgements

We thank Dan Stinebring and Maura McLaughlin for access to ISS data
from Arecibo Observatory, Don Backer for access to pulsar timing data
from the Green Bank 85~ft telescope, and Zaven Arzoumanian and Andrea
Lommen for useful discussions.  S.C. gratefully acknowledges the
hospitality and assistance of the National Radio Astronomy
Observatory, where part of this work was done. NRAO is a facility of
the National Science Foundation (NSF) operated under cooperative
agreement by Associated Universities, Inc. This work at Cornell was
supported in part by NSF grant AST 9819931, and by the National
Astronomy and Ionosphere Center, which operates Arecibo Observatory
under a cooperative agreement with the NSF. Basic research in radio
astronomy at the Naval Research Laboratory is supported by the Office
of Naval Research.


\end{document}